\documentclass{emulateapj}
\usepackage{graphicx,color}
\usepackage{amssymb,amsmath}
\usepackage[colorlinks,hyperfootnotes=false,citecolor=blue,linkcolor=blue]{hyperref}

\begin{document}
\shorttitle{A Survey for CIII] Emission at $\lowercase{z}\sim7-8$}
\shortauthors{Zitrin et al.}

\slugcomment{Submitted to the Astrophysical Journal}

\title{A Pilot Survey for CIII] Emission in the Reionization Era: Gravitationally-Lensed $\lowercase{z}\sim7-8$ Galaxies in the Frontier Fields Cluster Abell 2744}


\author{Adi Zitrin\altaffilmark{1,2}, Richard S. Ellis\altaffilmark{1}, Sirio Belli\altaffilmark{1}, Daniel P. Stark\altaffilmark{3}}

\altaffiltext{1}{Cahill Center for Astronomy and Astrophysics, California Institute of Technology, MC 249-17, Pasadena, CA 91125, USA; adizitrin@gmail.com}
\altaffiltext{2}{Hubble Fellow}
\altaffiltext{3}{Steward Observatory, University of Arizona, 933 N Cherry Ave, Tucson, AZ 85721 USA}


\begin{abstract}
We report results of a search for CIII] $\lambda \lambda$1907,1909 {\AA} emission using Keck's MOSFIRE spectrograph in a sample of 7 $z_{phot}\sim7-8$ candidates ($H\sim27$) lensed by the Hubble Frontier Field cluster Abell 2744. Earlier work has suggested the promise of using the CIII] doublet for redshift confirmation of galaxies in the reionization era given $Ly\alpha$ ($\lambda$1216 {\AA}) is likely attenuated by the neutral intergalactic medium. The primary challenge of this approach is the feasibility of locating CIII] emission without advanced knowledge of the spectroscopic redshift. With an integration time of 5 hours in the H-band, we reach a $5\sigma$ median flux limit (in between the skylines) of $1.5\times10^{-18}$ ergs cm$^{-2}$ sec$^{-1}$ but no convincing CIII] emission was found. We also incorporate preliminary measurements from two other CLASH/HFF clusters in which, similarly, no line was detected, but these were observed to lesser depth. Using the known distribution of OH emission and the photometric redshift likelihood distribution of each lensed candidate, we present statistical upper limits on the mean total CIII] rest-frame equivalent width for our $z\simeq7-8$ sample. For a signal/noise ratio of 5, we estimate the typical CIII] doublet rest-frame equivalent width is, with 95\% confidence, $<26\pm5$ {\AA}. Although consistent with the strength of earlier detections in brighter objects at $z\simeq6-7$, our study illustrates the necessity of studying more luminous or strongly-lensed examples prior to the launch of the James Webb Space Telescope.
\end{abstract}



\section{Introduction}\label{intro}
The reionization of the intergalactic medium (IGM) represents a key phase in the evolution of the Universe. Observations of high-redshift galaxies, which have charted a marked decline in the visibility of $Ly\alpha$ emission with redshift \citep{Stark2010z3-7fractions,Schenker2012DeclinLyAinZ}, and those of $z>5$ quasars which trace the redshift-dependent Gunn-Peterson absorption \citep[][]{Fan2006Quasars}, indicate that cosmic reionization was largely complete by $z\sim6$. The duration of the reionization process is constrained by the polarization of the microwave background due to Thomson scattering by electrons in the ionized era; recent data from the Planck satellite and results derived from the abundance and luminosity distribution of the $z>6$ galaxy population now suggest reionization was a rapid process which extended over $6 < z < 10$ \citep{Planck15,Robertson15}.

While $Ly\alpha$ emission ($\lambda$ 1216 {\AA}) has proven to be the most valuable spectroscopic indicator for faint star-forming galaxies in the redshift range $4<z<6$ \citep[e.g.][]{Stark2010z3-7fractions, Stark2011LAE}, resonant scattering by neutral gas in the IGM likely renders this line ineffective as a reliable probe beyond $z\simeq$6.5. Despite much observational effort, there are currently very few convincing cases of detected $Ly\alpha$ emission beyond $z\simeq7$ \citep{Ono2012,Finkelstein2013,Vanzella2014z6p4M0717,Schenker2014above7Spec,Oesch15} and several distant star-forming galaxies reveal no emission despite heroic exposure times \citep{Vanzella2014VLT52Hours}. As a result, \citet{Stark2014z2CIIILymanalphaZ2} proposed it may be feasible to use metallic lines in the ultraviolet (UV) as alternative spectroscopic indicators. Examining the spectra of 17 gravitationally-lensed low-luminosity galaxies at $z\simeq1.5-3$, they discuss the feasibility of searching for CIII] ($\lambda \lambda$1907,1909 {\AA}) and CIV ($\lambda \lambda$1548,1550 {\AA}) emission. Although such metallic lines are normally much weaker than $Ly\alpha$ in luminous systems, in young metal-poor low luminosity systems characteristic of those at high redshifts these lines may become relatively more prominent. In their sample of 17 $z\simeq1.5-3$ galaxies, \citet{Stark2014z2CIIILymanalphaZ2} find CIII] emission has an equivalent width (EW) which correlates with that of $Ly\alpha$ and is typically 10 times weaker. As an encouraging proof of concept, \citet{Stark2014CIIIdetectionz67} recently claimed tentative detections of CIII] emission in two $J\sim25.2$ galaxies with pre-determined $Ly\alpha$ emission at redshifts of $z=6.03$ and $z=7.21$.

However, as emphasized by Stark and collaborators, detecting CIII] emission in galaxies where its expected wavelength is {\it a priori} known from a $Ly\alpha$ redshift is less challenging than searching for emission across a wider range of wavelength governed only by a photometric redshift likelihood distribution, and given the density of skylines. Motivated by the interest in exploring the potential of this, possibly the only, immediate route to spectroscopic progress in the reionization era, we have embarked on a statistical search. Our plan is examine the spectra of a sample of gravitationally-lensed sources in the redshift range $z\sim$6.7-8.5 derived from recent compilations in several massive clusters \citep[e.g.][]{Bradley2013highz,Atek2014A2744,Zheng2014A2744,Coe2014FF}. Such a statistical approach is now possible due to the arrival of multi-slit near-infrared spectrographs such as Keck's \emph{Multi-Object Spectrometer For Infra-Red Exploration} (MOSFIRE; \citealt{KeckMOSFIRERef}). In this paper, we examine the practicality of the method with realistic exposure times for the Frontier Field cluster Abell 2744, and also include preliminary data for two other CLASH/HFF clusters observed to shallower depths. 

\begin{figure}
\centering
\includegraphics[width=0.5\textwidth,trim=1.2cm 0cm 1.0cm 0cm, clip=true]{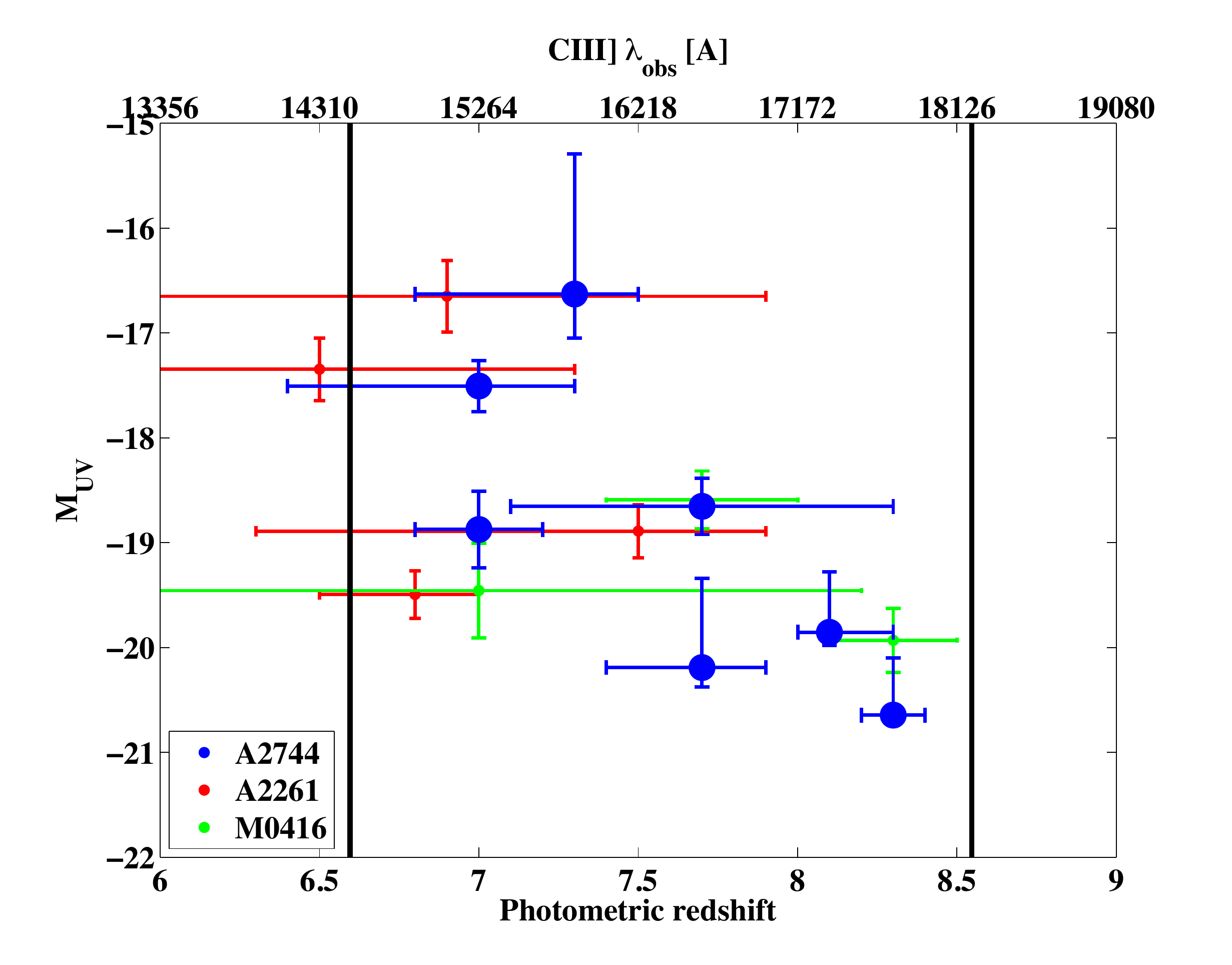}
\caption{The distribution of our candidate lensed galaxies in photometric redshift and UV absolute magnitude for A2744 and the two other clusters. Black lines denote the window within which CIII] would be visible at a wavelength indicated on the top axis.}\label{fig1}
\end{figure}

The paper is organized as follows: In \S \ref{obs} we overview the sample, observations, and data reduction. In \S \ref{results} we discuss the results, summarized in \ref{summary}. Throughout the work we use a standard $\Lambda$CDM cosmology with ($\Omega_{\rm m0}=0.3$, $\Omega_{\Lambda 0}=0.7$, $H_{0}=100$ $h$ km s$^{-1}$Mpc$^{-1}$, with $h=0.7$), and magnitudes are given using the AB convention. Cluster names are abbreviated to ``A'' or ``M'' for Abell (e.g. \citealt{Abell1989cat}) and MACS (MAssive Cluster Survey; e.g. \citealt{Ebeling2010FinalMACS}) clusters, respectively, followed by their ID.

\section{Data} \label{obs}

\begin{figure*}
\centering
\includegraphics[width=0.9\textwidth]{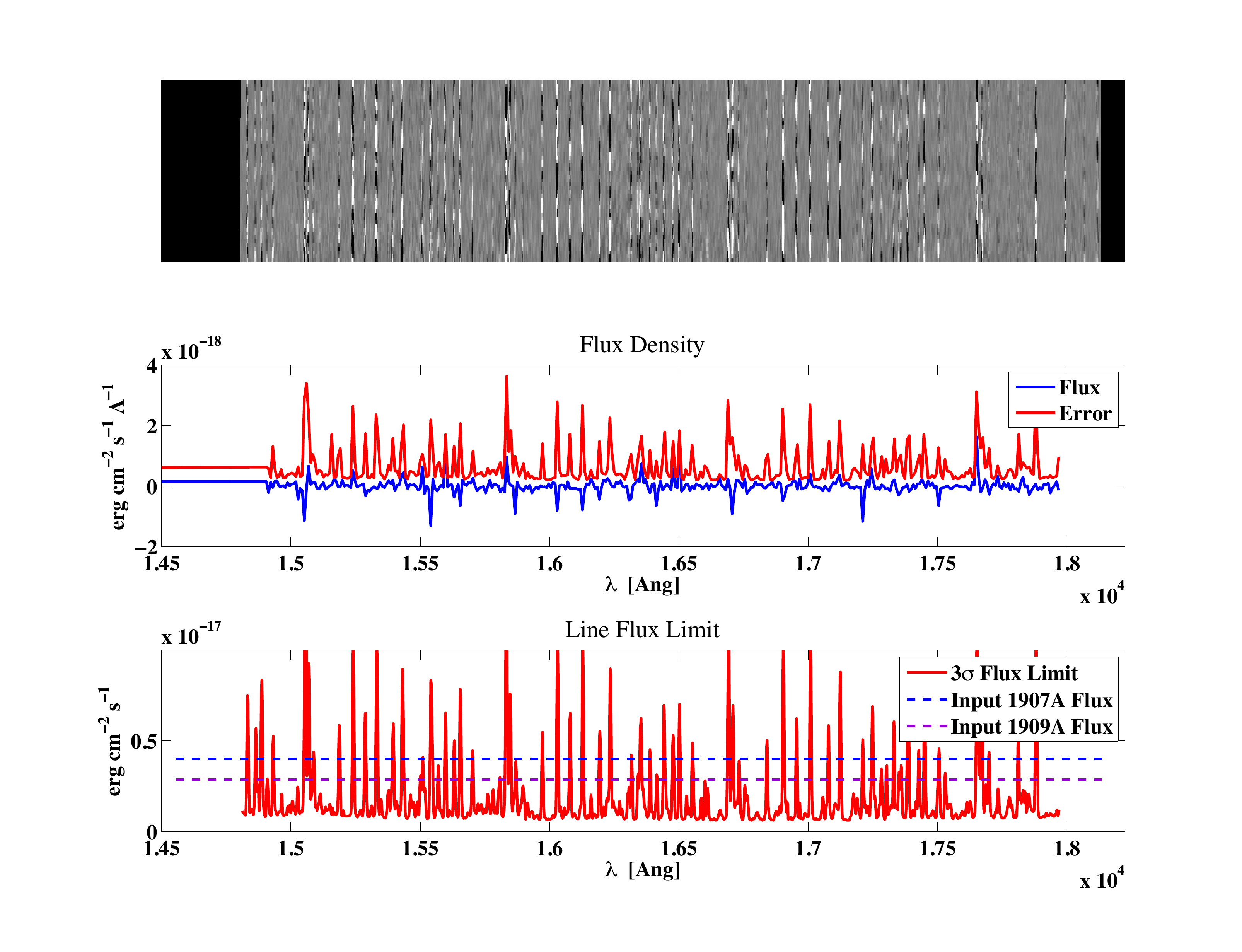}
\caption{Illustration of the data. The \emph{top} panel shows an arbitrary reduced slit in A2744, centered vertically on the CIII] candidate. The \emph{middle} panel shows the extracted 1D spectrum (\emph{blue}), and its $1\sigma$ error (\emph{red}) -- both smoothed here for illustrative purposes. The \emph{bottom} panel shows an example of a step in the procedure for the determining the upper-limit typical CIII] line flux in our sample. In \emph{red} we show the $3\sigma$ line flux limit, and the \emph{dashed blue and purple} horizontal lines show fiducial input CIII] line fluxes. For each such iteration we measure the fraction (in wavelength) in which the input flux is higher than the observational limit, weighted by the photometric redshift distribution. This indicates the chance of seeing CIII], as elaborated in \S \ref{results}.}\label{fig2}
\end{figure*}

\subsection{Target Selection}

We constructed a sample of high-$z$ candidates magnified by the galaxy cluster A2744 using photometry from Hubble Frontier Fields program (\citealt{Lotz2014AAS_FFreview}). Candidates for inclusion in our multi-slit mask were derived from \citet{Zheng2014A2744}, \citet{Coe2014FF} and \citet[][see also \citealt{Ishigaki2014}]{Atek2014A2744,Atek2014LF2}. We pre-selected targets of known magnification down to an apparent magnitude of $H_{160}\sim28$ with photometric redshifts in the range $6.7<z<8.5$, corresponding to the visibility of CIII] within MOSFIRE's $H$-band filter. We also included the $z\sim9.8$ candidate from \citet{Zitrin2014highz} to explore the option of detecting the CIV $\lambda\lambda$ (1548, 1550) \AA\ doublet in the same band. This follows a promising detection of CIV emission at $z\simeq$7.05 by \citet{Stark2015S}.  Photometric redshift likelihood distributions, $P(z)$, were obtained with the Bayesian Photometric Redshift code (BPZ; \citealt{Benitez2004,Coe2006}; see also \citealt{Zheng2014A2744}), covering the full redshift range available to HST (from $z=0.01$ to $z=12$ in $\Delta z=0.001$ increments). Due to slit-mask positioning constraints, only a subset of good candidates could be included on the MOSFIRE mask and priority was given to brighter galaxies. The final mask included 8 high-$z$ candidates as summarized in Table \ref{galaxiesTable}. The CIII] candidates lie mainly in the magnitude range $26.2<H<27.5$ and have photo-z uncertainties of $\delta\,z\simeq0.2-0.3$. The rest-frame UV luminosities corrected for lensing magnifications have a mean of $M_{UV}\simeq-18.8$ and standard deviation of 1.4. As part of this campaign we also have begun observations of two further lensing clusters, MACS0416 and A2261, drawing candidates from the catalogs of \citet{Bradley2013highz} and \citet[][see also \citealt{McLeod2014z9}]{Coe2014FF}. As the photometric redshift distributions of these galaxies are somewhat less secure (especially for A2261), and since our observations of these clusters are significantly shallower, they currently provide less useful constraints on the presence of CIII], although we incorporate the results in this paper. We also list these additional sources in Table \ref{galaxiesTable}. Fig. \ref{fig1} summarizes the redshift and UV luminosity distribution of the total sample in the context of the $H$-band window available for detecting CIII] with MOSFIRE.

\begin{figure*}
\centering
\includegraphics[width=0.47\textwidth,height=0.47\textwidth,trim=1.8cm 0cm 1.8cm 0cm, clip=true]{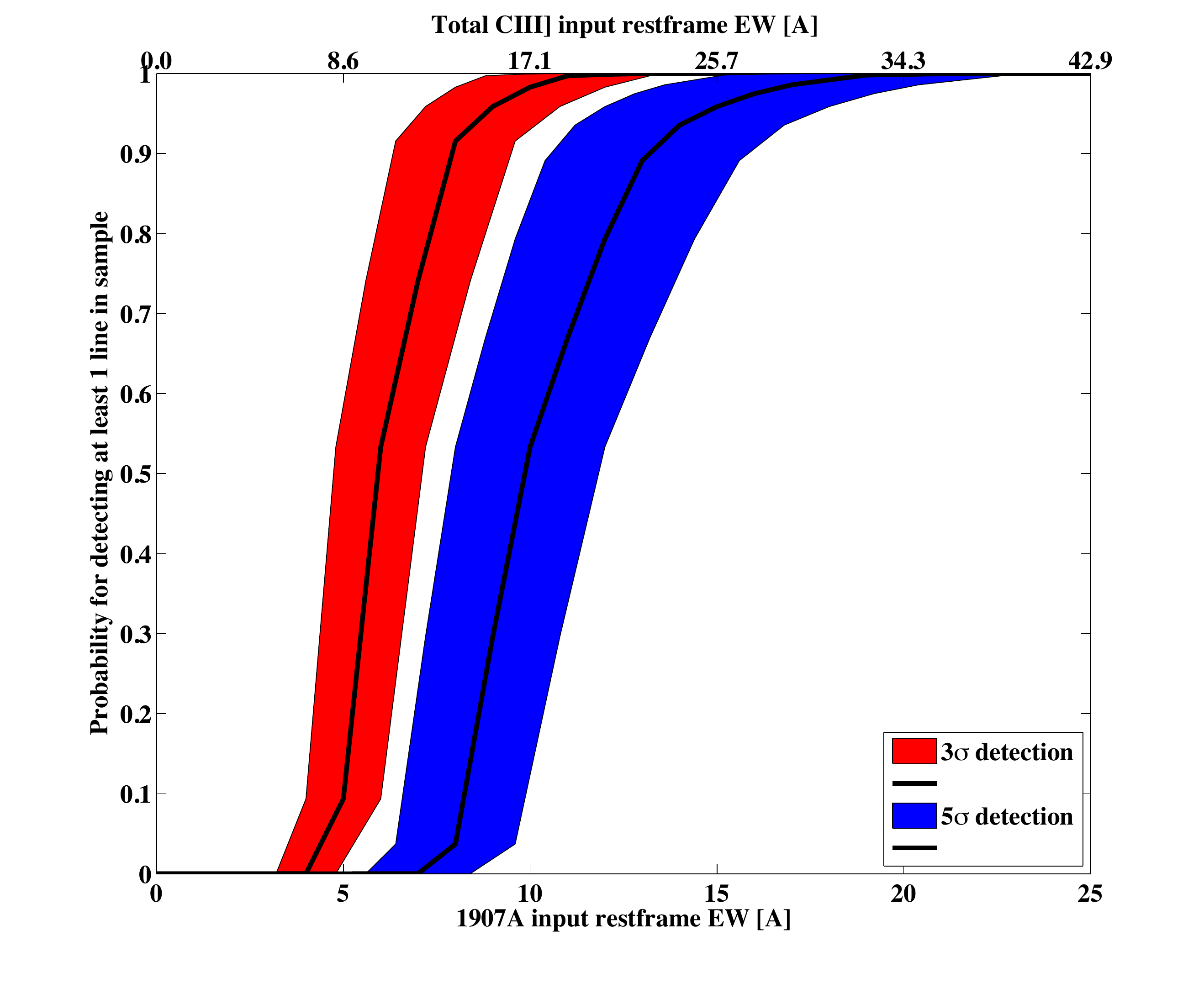}
\includegraphics[width=0.47\textwidth,height=0.47\textwidth,trim=1.8cm 0cm 1.8cm 0cm, clip=true]{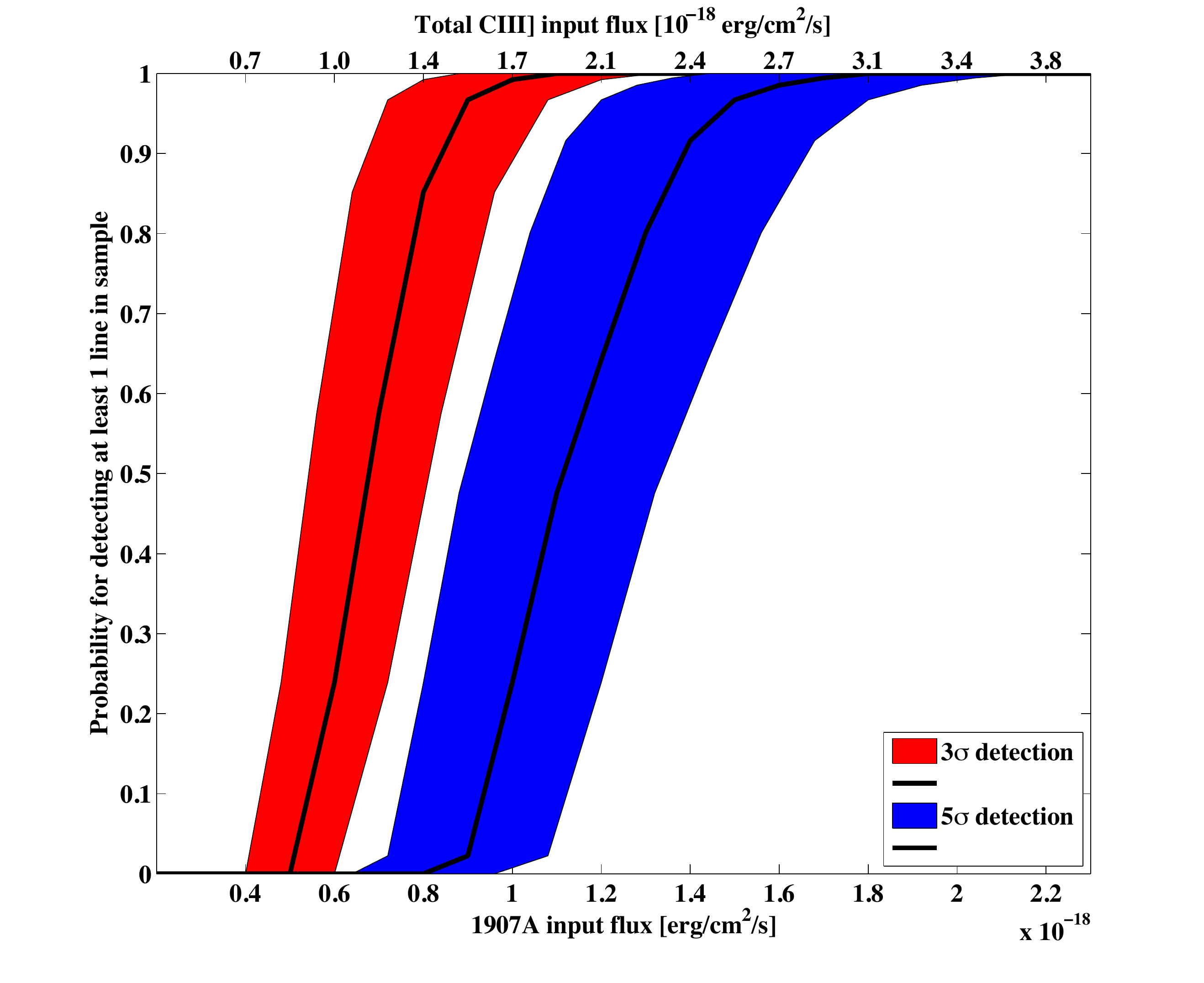}\\
\includegraphics[width=0.47\textwidth,height=0.47\textwidth,trim=1.8cm 0cm 1.8cm 0cm, clip=true]{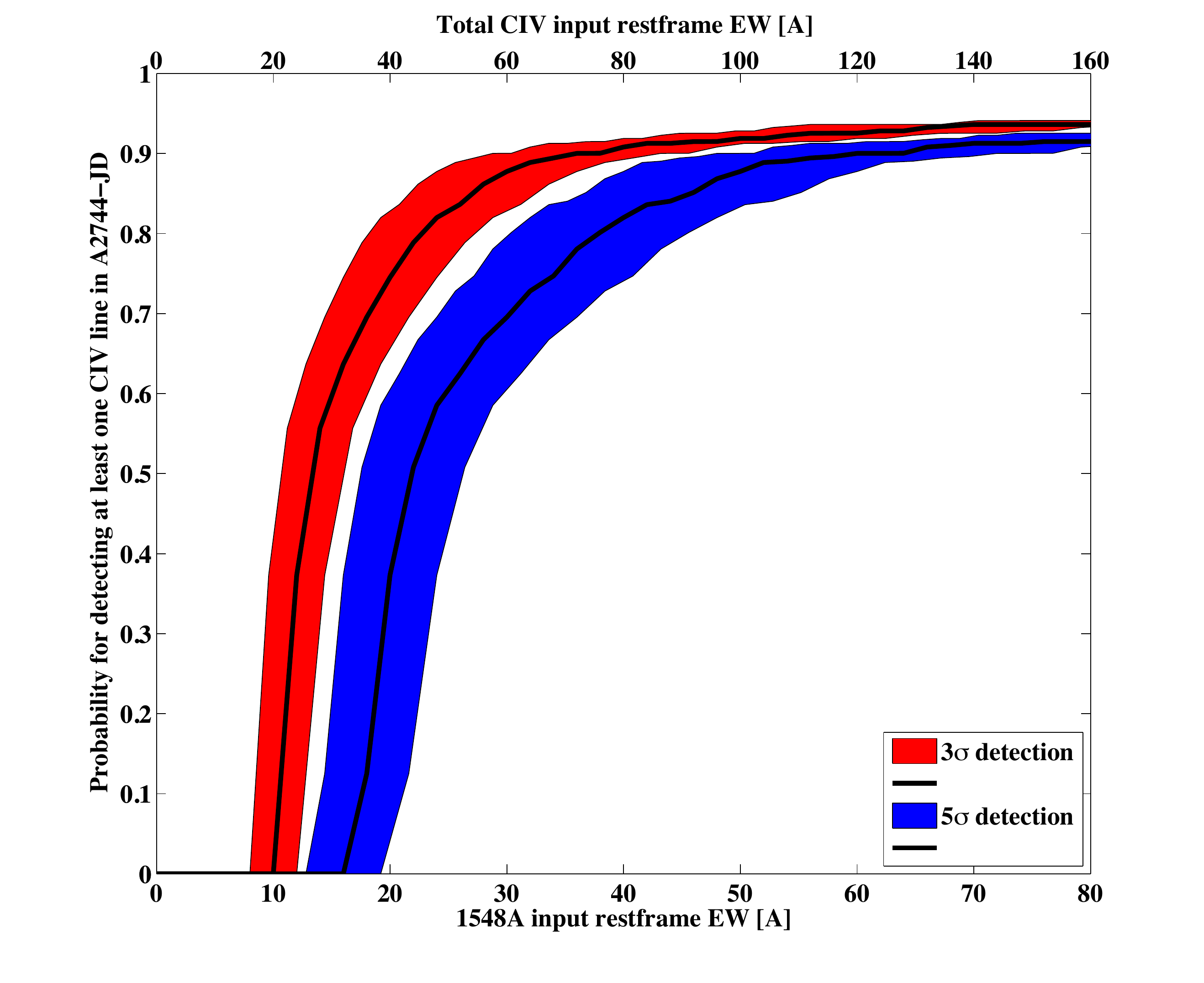}
\includegraphics[width=0.47\textwidth,height=0.47\textwidth,trim=1.8cm 0cm 1.8cm 0cm, clip=true]{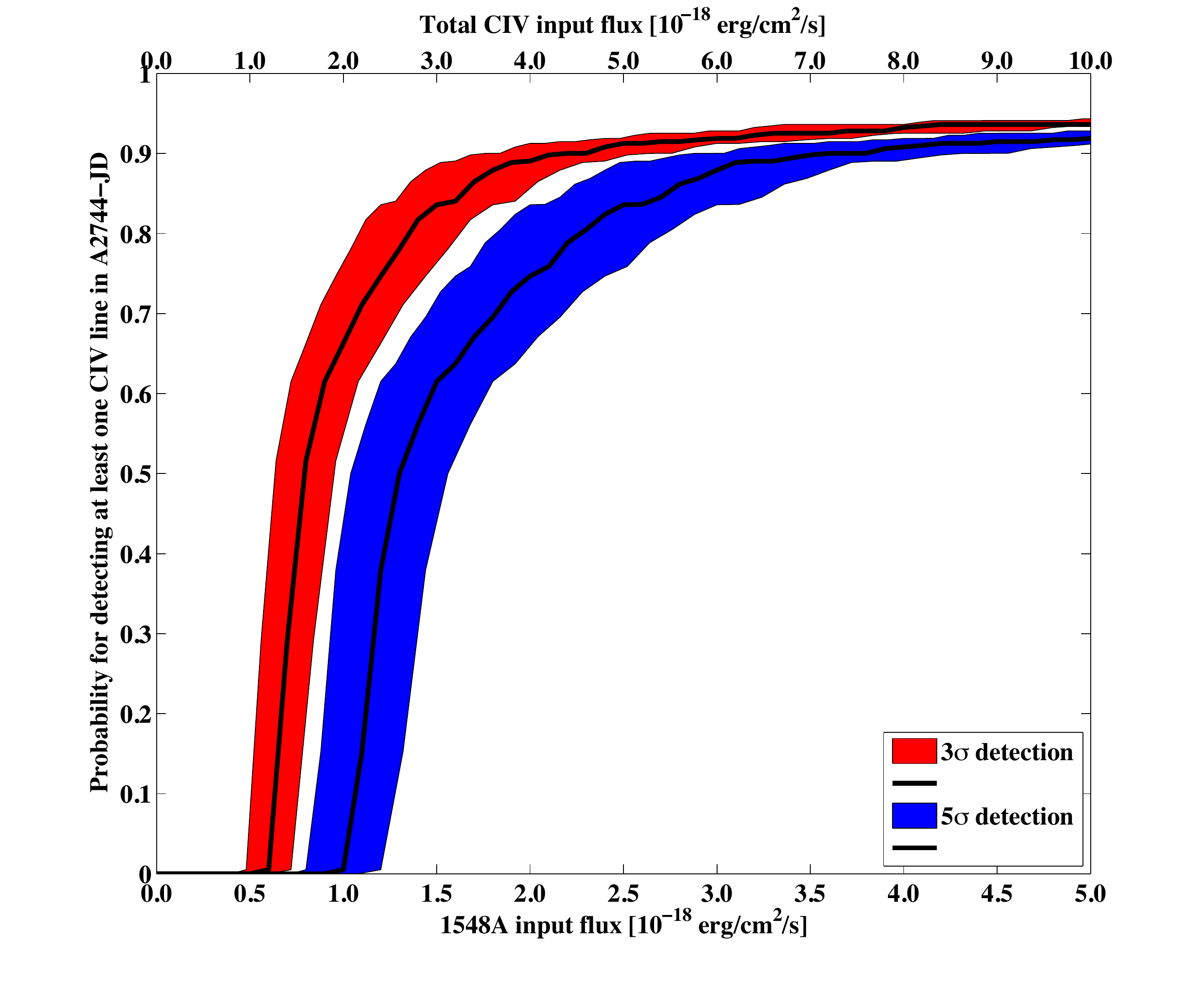}
\caption{\emph{Top Left:} Probability to detect at least one CIII] line in our sample, as a function of the total (top axis) and $1907$ {\AA} (bottom axis) CIII] rest-frame equivalent width. We plot the probabilities for detection thresholds of 3, and 5$\sigma$, and the shaded regions indicate the errors based on our adopted 20\% flux calibration error. \emph{Top Right:} A similar plot in terms of the total and $1907$ {\AA} line flux.\\
\emph{Bottom:} Probability to detect at least one of the two CIV lines in the $z\sim9.8$ object \citep{Zitrin2014highz}, as a function of line flux and restframe EW.}\label{fig3}
\end{figure*}

\subsection{Observations and Data Reduction}

Observations with MOSFIRE on the Keck 1 telescope were undertaken on 16 September, 25 November, and 27 November 2014. A total exposure of 5 hours was obtained for A2744. Thus far only 2.7 and 2.2 hours have been secured for A2261 and M0416, respectively. Median seeing varied between $\sim0.5-0.8\arcsec$. Each exposure comprised 120 sec integrations with an AB dithering pattern of $\pm1.25\arcsec$ along the slit. On each mask, one slit was assigned to an alignment star, in order to track possible positional drifts and transparency changes.

Data reduction was performed using the standard MOSFIRE reduction pipeline\footnotemark[4] \footnotetext[4]{http://www2.keck.hawaii.edu/inst/mosfire/drp.html}. For each flat-fielded slit we extracted the 1D spectrum using a 11 pixel boxcar centered on the expected position of the target. A similar procedure was adopted in quadrature to derive the 1$\sigma$ error distribution. The addition of data from different nights was performed by inverse-variance averaging the calibrated 1D spectra. To obtain the $1\sigma$ flux limits for CIII], we assumed a marginally resolved line width \citep{Stark2014CIIIdetectionz67} corresponding to three MOSFIRE pixels ($\sim5$ {\AA} in the $H$-band) and summing in quadrature the 1D $\sigma$ spectrum within the 3-pixel window.

Spectrophotometric standard stars were observed twice a night in similarly good conditions.  We scaled a Vega model\footnotemark[5] \footnotetext[5]{http://kurucz.harvard.edu/stars/vega/} to each standard star to determine a wavelength-dependent flux calibration using the procedure described in \citet{Vacca2003calibration}, which more accurately traces telluric corrections free from stellar absorption features. Our various flux calibrations on each night agree to within $15\%$ and are also consistent with the nominal MOSFIRE calibration files (C. Steidel, \emph{private communication}) to within $\sim10\%$. The calibrated spectra of alignment stars (incorporated on our multi-slit mask) from different nights also agree to within $1-8\%$, and are typically within $\sim20\%$ of the flux level expected from their H$_{160}$ photometry, after aperture corrections for slit losses. Using our adopted calibration, the median $5\sigma$ detection limit achieved in between the OH skylines for our A2744 exposure is $1.5\times10^{-18}$ cgs. This line flux limit is comparable with that achieved by \citet{Stark2014CIIIdetectionz67} using MOSFIRE. In their 3.1 hour exposure, they report a limiting flux (5$\sigma$) of $1.8\times10^{-18}$ cgs.

\section{Results} \label{results}

All reduced spectra were visually inspected given the expected wavelength range where CIII] might be visible according to the photometric redshift likelihood function. No convincing line was seen for any of the 7 CIII] candidates. We thus seek to determine the likely range of fluxes and equivalent widths (EW) for CIII] consistent with our non-detections\footnotemark[6]\footnotetext[6]{We note that A2744 was also observed with the same mask for 2 hours in the Y-band, searching for Ly$_{\alpha}$. No line was detected and these data will be presented elsewhere.}. In other words, we estimate the probability of detecting at least one CIII] line in our survey, as a function of a given mean total flux and EW. Since CIII] is a doublet, we assume a line ratio CIII] 1907/1909 {\AA} of 1.4 \citep{Stark2014CIIIdetectionz67}. We consider total line strengths in the range $0-4\times10^{-18}$ ergs cm$^{-2}$ sec$^{-1}$ in $5\times10^{-20}$ ergs cm$^{-2}$ sec$^{-1}$ increments as illustrated in Fig. \ref{fig2}. For each doublet line, and for each redshift step, we checked if its input flux would exceed a certain detection significance ($x\sigma$) in the corresponding wavelength in the observer frame, where $x$ is a chosen signal/noise ratio. For a fixed line flux, the likelihood of having a detection with $x\sigma_{k}$ significance in the examined slit $k$ is given by the number of spectral pixels with positive detections over the total number of pixels, weighted by the redshift probability function $P(z)$. The relevant expression for slit $k$ can be formulated as:
\begin{equation}\label{chances1}
P_{{det,k}}(F_{in},x)=
\frac{\sum_{i}P_{k}(z_{i}) \Theta\left(F_{in},x,\sigma_{k},z_i\right)} {\sum_{i}P_{k}(z_{i})} \; ,
\end{equation}
where the sum is over all the redshift steps $z_i$ ($z=0$ to $z=12$ in 0.001 increments), and $\Theta$ is defined as
\begin{displaymath}
   \Theta \left(F_{in},x,\sigma_{k},z_i\right) = \left\{
     \begin{array}{lr}
       1 & \mathrm{if\ } F_{in} > x\sigma_k(1907\mathrm{\AA} \cdot (1+z_i)) \\
         & \mathrm{\ or\ } F_{in}/1.4 > x\sigma_k(1909\mathrm{\AA} \cdot (1+z_i)) \\
       0 & \mathrm{otherwise,}
     \end{array}
   \right.
\end{displaymath}
and we define $\sigma_k=\infty$ for $z_i$ steps placing the line outside the MOSFIRE $H$ band.
$P_{det,k}$ therefore provides a \emph{conditional} probability, i.e. the chance of detecting \emph{at least} one of the two CIII] lines for a given target $k$ with redshift probability distribution $P_{k}(z)$, given the limiting noise in our spectra for the mask, $\sigma_{k}(\lambda)$ (see \S2), and as a function of the input line flux $F_{in}$ and the detection significance $x$. Finally, the probability of detecting at least one line of a given flux $F_{in}$ over the entire sample is:
\begin{equation}\label{chances1Global}
P_{sample}(F_{in},x)=1-\prod_{k}\left(1-P_{det,k}(F_{in},x)\right),
\end{equation}
where the product is over all slits.

We repeat the above process also in terms of restframe EW, where in each iteration, instead of running over a range of input fluxes, we run over a range of restframe EWs, translated in each iteration, for each object individually, to the corresponding input flux.

Figure \ref{fig3} shows the probability, for both 3 and 5 $\sigma$ detections, of finding at least one line in our A2744 survey as a function of the mean total CIII] flux and rest-frame EW. For example we have a 95\% chance of detecting at least one CIII] line in our MOSFIRE survey at 5$\sigma$ significance, if the typical CIII] $\lambda$1907 \AA\ line flux is $\simeq1.5 \times 10^{-18}$ ergs cm$^{-2}$ sec$^{-1}$ (total doublet flux of $\simeq2.6 \times 10^{-18}$ ergs cm$^{-2}$ sec$^{-1}$), or equivalently, if the rest-frame EW for the combined CIII] doublet is $26\pm5$ \AA\ or higher. In this estimate we have included limits from the shallower exposures on A2261 and M0416, but the results remain similar (to within typically 5\%) if the sample is restricted to A2744 -- for which photometric redshift errors are typically smaller and the observations are significantly deeper. Errors were propagated assuming our adopted $20\%$ uncertainty in the flux calibration.

For comparison, \citet{Stark2014CIIIdetectionz67} detected, with $3.3\sigma$, a $\lambda$1909 \AA\ CIII] line of $\simeq4.2\pm1.2 \times 10^{-18}$ ergs cm$^{-2}$ sec$^{-1}$ (total estimated CIII] flux $\simeq1.1\pm0.3 \times 10^{-17}$) in a $z=6.03$ galaxy ($J=25.2$), and a $2.8\sigma$ CIII] detection of likely $\lambda$1909 \AA\ of $\simeq0.9\pm0.3 \times 10^{-18}$ ergs cm$^{-2}$ sec$^{-1}$ (total estimated CIII] flux $\simeq2.3\pm0.5 \times 10^{-18}$) in a $z=7.21$ galaxy ($J=25.2$). The total CIII] restframe EWs of these detections are $22.5\pm7.1$ \AA\ and $7.6\pm2.8$ \AA\, respectively. 

\begin{figure}
\centering
\includegraphics[width=0.5\textwidth,trim=1.3cm 0cm 1.1cm 0cm, clip=true]{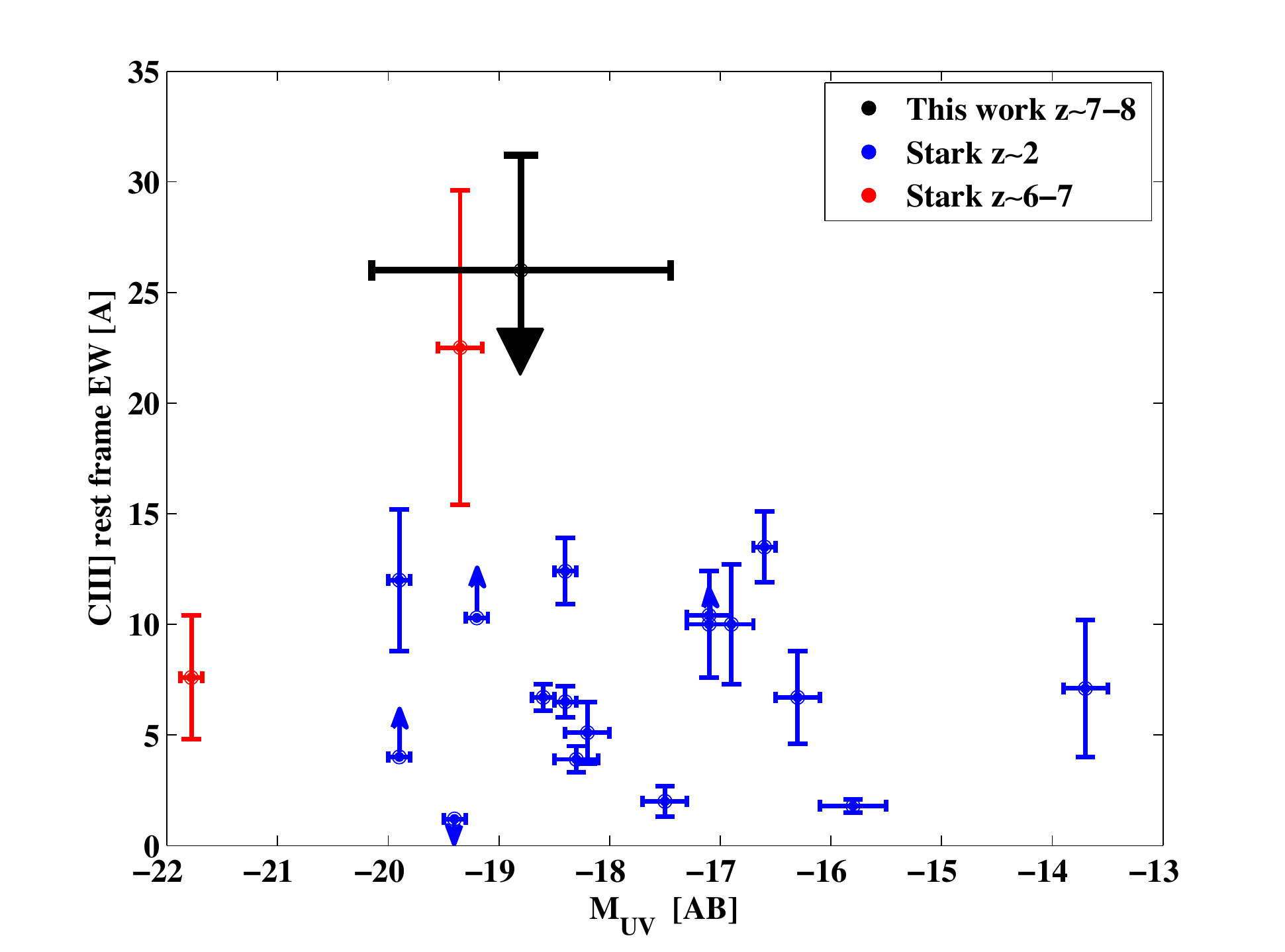}
\caption{The rest-frame EW as a function of absolute UV magnitude, of previous measurements of CIII] from \citet{Stark2014z2CIIILymanalphaZ2,Stark2014CIIIdetectionz67}. The \emph{black} error bar and arrow shows the (95\% C.L.) limit obtained in this work. }\label{fig5}
\end{figure}

While our observational limits are deep enough to recover similar line fluxes to those found by \citet[][see also \citealt{Stark2014z2CIIILymanalphaZ2}]{Stark2014CIIIdetectionz67}, our galaxies are fainter ($\sim27$ AB), so that our limits on the rest-frame EWs are less constraining. Assuming a CIII] EW of 22.5(7.6) \AA\ as was found by Stark, we have $\sim99.9\%(10\%)$ chance for detecting at least one such line in our sample with 3$\sigma$, or $\sim90\%(0\%)$ for 5$\sigma$. Thus, it is quite likely that the primary reason for the non-detection in our survey is that, on average, the present sample is significantly fainter than those targeted by Stark et al., which also had the benefit of secure Ly$\alpha$-based redshifts. The main conclusion of our limits seen in Figures \ref{fig3} and \ref{fig5} is that even with a more ambitious spectroscopic campaign that would likely increase the exposure time by a factor $\times$4 (corresponding to a 3 night integration on one mask), only more luminous $z\simeq$7-8 galaxies in the reionization era would appear to be amenable for study with any reliability. Alternatively, brighter and/or more highly-magnified examples, such as those close to the critical line of a foreground cluster, might provide promising targets although generally such sources are rare. It is interesting to note the non-detection (and upper limit) on CIII] emission recently claimed by \citet{Darach2015LyA1689} for a brighter source with $H$=24.7, magnified by $\mu\sim10$ at $z$=7.5, showing that even for significantly brighter objects CIII] detection can be challenging. Searching for bright magnified dropouts in a very large sample of clusters, for example, is desirable for progress with current facilities and might help deliver JWST with first light targets.

\vspace{0.2cm}

For completeness, we also calculate the limit on the CIV $\lambda \lambda$ (1548,1550) {\AA} doublet for the $z\simeq9.8$ multiply-imaged object discovered by \citet{Zitrin2014highz} behind A2744 (Fig. \ref{fig3}). \citet{Stark2014z2CIIILymanalphaZ2} found prominent CIV emission in some of the $z\sim2$ galaxies they targeted, and highlighted CIV as an additional promising diagnostic for high-redshift galaxies. Typically they found CIV line fluxes only a factor of about 2 weaker than those of CIII]. At the proposed redshift of the \citet{Zitrin2014highz} object, the doublet would be readily resolved and we assume both lines have equal strength. In this case, we determine that, with $\sim90\%$ confidence, the line flux for either one of the two CIV lines for a detection significance of 5$\sigma$ is less than $\simeq3.6\times10^{-18}$ ergs cm$^{-2}$ sec$^{-1}$, and the rest-frame EW less than $\simeq32$ {\AA}. This translates to a magnification-corrected, total CIV luminosity of $\lesssim3.9\times10^{41}$ erg sec$^{-1}$ at $z=9.8$. It would be interesting to investigate further the properties of CIV emission in a larger sample. Note also \citet{Stark2015S} have now detected a promising a CIV $\lambda1548$ {\AA} line in a $z\simeq7.05$ object, corresponding to a restframe EW of $\simeq18.1$ {\AA} (a total CIV restframe EW of 38 {\AA}). As in the CIII] case, despite reaching deep enough to detect a similar line flux, $\simeq4.1\times10^{-18}$ ergs cm$^{-2}$ sec$^{-1}$, in terms of EW the non-detection is consistent with the limits obtained from the single $z\sim10$ object.


\medskip

\section{Conclusions} \label{summary}
Given the attenuation of Ly$\alpha$ by neutral gas in the reionization era, the CIII] doublet has been proposed as a promising route toward spectroscopic verification and study of high-redshift candidates \citep{Stark2014CIIIdetectionz67,Stark2014z2CIIILymanalphaZ2}. We report results from a short campaign with Keck/MOSFIRE to assess the prospects of detecting CIII] lines in a sample of faint gravitationally-lensed $z\sim7-8$ galaxies where $Ly\alpha$ is not seen and thus the search window in wavelength is much larger than in earlier work. We observed 14 high-$z$ candidates magnified by three galaxy clusters. For our deepest field (A2744, with 7 CIII] candidates), we reached a 5$\sigma$(3$\sigma$) flux limit of $1.5(0.9)\times10^{-18}$ ergs cm$^{-2}$ sec$^{-1}$ but did not detect any convincing line. Using a statistical method employing data from our collective campaign, we provide upper limits on the typical CIII] line flux and its rest-frame EW. Although our limits reach the line fluxes observed in some actual CIII] detections claimed in the recent literature, because our sample is significantly fainter in apparent magnitude, we only marginally reach the expected EWs based on these recent detections. This demonstrates the challenge of continuing the present investigation with current observing facilities
unless either (i) brighter or more strongly-lensed
sources are targeted and/or (ii) the CIII] is
found to be more prominent in intrinsically-fainter
systems \citep[e.g.][]{Stark2014z2CIIILymanalphaZ2}. More data is needed to test this latter suggestion.


\section*{acknowledgments}
We thank the reviewer of this work for important comments. AZ is grateful for data products generated by Dan Coe and the CLASH team (PI: M. Postman), and for data supplied to us by PI: Wei Zheng, Alberto Molino and group (HST grant AR 13279). AZ thanks Carrie Bridge and Adam Miller for helpful discussions, and Chuck Steidel for sharing their nominal calibration files. Support for this work was provided by NASA through Hubble Fellowship grant \#HST-HF2-51334.001-A awarded by STScI, which is operated by the Association of Universities for Research in Astronomy, Inc. under NASA contract NAS~5-26555. This work is in part based on previous observations made with the NASA/ESA Hubble Space Telescope. The data presented herein were obtained at the W.M. Keck Observatory. The authors wish to recognize and acknowledge the very significant cultural role and reverence that the summit of Mauna Kea has always had within the indigenous Hawaiian community. We are most fortunate to have the opportunity to conduct observations from this mountain.

\tabletypesize{\small}
\setlength{\tabcolsep}{0.02in}
\begin{deluxetable*}{|c|c|c|c|c|c|c|c|}
\tablecolumns{8}
\tablewidth{1\textwidth}
\tablecaption{\small{The Sample} \label{galaxiesTable}}
\tablehead{
\colhead{ID
} &
\colhead{$\alpha$(deg.)
} &
\colhead{$\delta$(deg.)
} &
\colhead{Phot-$z$
} &
\colhead{$H_{160}$
} &
\colhead{$\mu$
} &
\colhead{$\beta$
} &
\colhead{$M_{UV,1500}$
}
}
\startdata
A2744-YD7$^{a,b}$ &  3.603397 & -30.382256 & $8.3^{+0.1}_{-0.1}$ &  $26.17\pm0.03$ & $1.4^{+0.7}_{-0.1}$  & $-1.38\pm1.86$  &   $-20.65_{-0.08}^{+0.54}$\\
A2744-ZD3$^{a,b,c}$  & 3.606477 & -30.380993 & $7.7^{+0.2}_{-0.3}$ & $26.45\pm0.04$ & $1.3^{+1.0}_{-0.1}$ & $-1.14\pm0.26$  &   $-20.19_{-0.19}^{+0.85}$\\
A2744-ZD9$^{a}$  & 3.603208 & -30.410368 & $7.0^{+0.2}_{-0.2}$ &  $26.48\pm0.04$ & $3.4^{+0.8}_{-0.8}$ & $-1.17\pm0.23$  &   $-18.88_{-0.37}^{+0.37}$\\
A2744-ZD7A2$^{a}$ & 3.592160 & -30.409925 & $7.3^{+0.2}_{-0.5}$ &  $28.18\pm0.04$ & $6.4^{+7.8}_{-2.2}$  & $-1.29\pm1.22$  &   $-16.63_{-0.42}^{+1.34}$\\
A2744-YD8$^{a,b,c}$ & 3.596096 & -30.385832 & $8.1^{+0.2}_{-0.1}$ & $26.65\pm0.04$ & $1.9^{+1.0}_{-0.2}$ & $-1.84\pm1.64$  &   $-19.86_{-0.13}^{+0.57}$\\
Atek-3772$^{c}$ & 3.5978343 & -30.395960 & $7.0^{+0.3}_{-0.6}$ & $27.45\pm0.05$ & $\sim$6.8 & $-1.77\pm1.00$  &   $-17.51_{-0.24}^{+0.24}$\\
Atek-5918$^{c,e}$ & 3.5951375 & -30.381131 & $7.7^{+0.6}_{-0.6}$ & $26.92\pm0.02$ & $\sim$3.5 & $-1.07\pm0.19$  &   $-18.65_{-0.27}^{+0.27}$\\
A2744-JDB$^{d}$ & 3.5950200  & -30.400750 &  $9.8^{+0.2}_{-0.4}$ &  $27.30\pm0.07$ & $11.3^{+4.8}_{-2.5}$& \nodata &  $\sim-17.6$ \\
\hline
A2261-0450$^{f}$ &  260.6124593 & 32.1438429 & $6.8^{+0.2}_{-0.3}$ & $25.5\pm0.06$ & $\sim5.6$ & $-1.85\pm0.15$  &   $-19.50_{-0.23}^{+0.23}$\\
A2261-0731$^{f}$ & 260.6232556 & 32.1393984 & $6.9^{+1.0}_{-5.9}$  & $27.9\pm0.22$ & $\sim7.7$ & $-1.00\pm0.67$  &   $-16.65_{-0.34}^{+0.34}$\\
A2261-0772$^{f}$ & 260.6059024 & 32.1388049 & $6.5^{+0.8}_{-5.4}$ & $27.4\pm0.19$  & $\sim6.3$ & $-2.17\pm0.64$  &   $-17.35_{-0.30}^{+0.30}$\\
A2261-0187$^{f}$ & 260.6073833 & 32.1495175 & $7.5^{+0.4}_{-1.2}$ & $27.0\pm0.13$ & $\sim2.9$ & $-1.18\pm1.47$  &   $-18.89_{-0.25}^{+0.25}$\\
\hline
MACS0416-0036$^{f}$ & 64.0260447 & -24.0509958 & $7.0^{+1.2}_{-6.0}$ & $26.8\pm0.16$ & $\sim1.3$ & $-0.56\pm0.45$  &   $-19.46_{-0.45}^{+0.45}$\\
Zheng-4408$^{g}$ &  64.0603330 & -24.064960  & $7.7^{+0.3}_{-0.3}$ & $27.85\pm0.08$ & $2.2^{+0.3}_{-0.3}$ & $-3.49\pm1.33$  &   $-18.59_{-0.28}^{+0.28}$\\
FFC2-1151-4540$^{b,g}$ & 64.0479780 & -24.081678 & $8.3^{+0.2}_{-0.2}$ & $26.59\pm0.03$ & $1.8^{+0.5}_{-0.5}$& $-1.44\pm0.69$  &  $-19.93_{-0.31}^{+0.31}$\\
\hline
\enddata
\tablecomments{\\
$\emph{Column 1:}$ Dropout's ID and references. The first work cited for each object represents the original source of photometric data and analysis, although in some cases we made adjustments to enhance consistency across the sample. \\
$\emph{Columns 2 \& 3:}$ RA and DEC in J2000.0. \\
$\emph{Column 4:}$ Photometric redshift and 95\% errors. \\
  $\emph{Column 5:}$ HST's apparent H$_{160}$-band magnitude. \\
  $\emph{Column 6:}$ Lensing magnification. If no error is listed a nominal $\sim20\%$ error is adopted \citep{Zitrin2015}.\\
   $\emph{Column 7:}$ UV-slope, $\beta$ ($1\sigma$ errors), calculated by a weighted least-squares fit.\\
   $\emph{Column 8:}$ Absolute magnitude, $M_{UV}$, at $\lambda=1500$ {\AA}, calculated from the said $F_{\lambda}\propto\lambda^{\beta}$ fit, where the error includes in quadrature the discrepancy from the absolute magnitude obtained by translating the flux in the WFC3 band containing the redshifted $\lambda=1500$ {\AA}, and the propagated photometric and magnification errors.\\}
\tablenotetext{a}{\citet{Zheng2014A2744}}
\tablenotetext{b}{\citet{Coe2014FF}}
\tablenotetext{c}{\citet{Atek2014A2744,Atek2014LF2}}
\tablenotetext{d}{\citet{Zitrin2014highz}. CIV target.}
\tablenotetext{e}{Our independent photo-z estimate permits a solution at $z\simeq$2}
\tablenotetext{f}{\citet{Bradley2013highz}}
\tablenotetext{g}{W. Zheng, private communication (\emph{in preparation}).}
    \end{deluxetable*}


\end{document}